\documentclass[prb,twocolumn,bibnotes,showpacs,amsbsy,floatfix,amsbsy,floatfix,superscriptaddress]{revtex4-2}
\usepackage{epsfig,color}
\usepackage[utf8]{inputenc} 
\usepackage[T1]{fontenc}    
\usepackage{amsmath}
\usepackage{amssymb}
\usepackage{amsfonts}
\usepackage{color}
\usepackage{wasysym}
\usepackage{latexsym} 
\usepackage[dvipsnames]{xcolor}
\usepackage{hyperref}
\hypersetup{
    colorlinks=true,       
    linkcolor=blue,        
    citecolor=blue,        
    filecolor=magenta,     
    urlcolor=blue          
}
\usepackage{graphicx,color}

\begin{document}

\title{Local and nonlocal STM transport signatures of spin polarization in second order topological superconductors}

\author{Pawe\l{} Szumniak}
\affiliation{AGH University of Science and Technology, Faculty of
Physics and Applied Computer Science,\\
al. Mickiewicza 30, 30-059 Krak\'ow, Poland}
\author{Daniel Loss}
\affiliation{Physics Department and Quantum Center, King Fahd University of Petroleum and Minerals, 31261, Dhahran, Saudi Arabia}
\affiliation{Department of Physics, University of Basel, Klingelbergstrasse 82, 4056 Basel, Switzerland}
\author{Jelena Klinovaja}
\affiliation{Department of Physics, University of Basel, Klingelbergstrasse 82, 4056 Basel, Switzerland}

\begin{abstract}
We investigate numerically the spin and transport properties of two-dimensional second-order topological superconductors (2D SOTSCs) hosting a pair of Majorana corner states (MCSs). First, we show that MCSs in the considered setup are characterized by a distinct spatial distribution of electronic spin polarization in the direction perpendicular to an applied in-plane magnetic field, with opposite signs for each MCS. Such a property can be used to label MCSs in a pair by their electronic spin. We propose a comprehensive spin-resolved transport protocol for measuring such a spin texture and further detecting the braiding (exchange) of a pair of MCSs, a crucial prerequisite for topological quantum computing. To be specific, we show that the magnitude of local conductance and the sign of nonlocal conductance are precisely linked to the sign of the electronic part of the MCS spin density and the spin polarization of the probe. Moreover, we show that the proposed technique can be used to detect the spin density profile of higher-energy quasiparticle states, e.g., edge states hosted in the SOTSC. We showed that all analyzed features are highly robust to strong static disorder , which makes our findings a clear experimental pathway to verify the spin structure of MCS and other quasiparticles hosted in SOTSCs.
\end{abstract}

\maketitle

\section{Introduction}

The topological phases of matter have been of central interest in condensed matter physics in recent years due to both their fundamental significance and practical potential applications, e.g., in fault-tolerant topological quantum computation (TQC)~\cite{freedman2002modular, freedman2002simulation, kitaev2003fault, nayak_non-abelian_2008}. Conventional $d$-dimensional topological insulators (TIs) and superconductors (TSCs) are characterized by a nontrivial gapped bulk that hosts topologically protected, gapless states on their $(d-1)$-dimensional boundaries~\cite{Hasan_Topological_2010, Qi_Topological_2011, bernevig2013topological, Chiu_Classification_2016}, which is a consequence of the bulk boundary correspondence. Recently, these concepts have been generalized into a more complete classification that covers so-called {\it high-order} topological states of matter \cite{benalcazar_quantized_2017, benalcazar_electric_2017, langbehn_reflection-symmetric_2017, parameswaran_topological_2017, song__2017, peng_boundary_2017, imhof_topolectrical-circuit_2018, schindler_higher-order_2018, geier_second-order_2018, serra-garcia_observation_2018, trifunovic_higher-order_2019, Slager_Impurity_2015, mittal_photonic_2019, xue_acoustic_2019, ni_observation_2019, el_hassan_corner_2019, rodriguez-vega_higher-order_2019, ghosh_higher_2020, Laubscher_Majorana_2020, Laubscher_Kramers_2020, Hirosawa_Magnonic_2020, ren_engineering_2020, szumniak_hinge_2020, xie_higher-order_2021, costa_discovery_2021, zhu_time-periodic_2022, yang_higher-order_2024, chatterjee_second-order_2024, han_cornertronics_2024, hussain_pentagonal_2024, luo_characterization_2025, qian_programmable_2025}. The characteristic new feature of such phases is that a $d$-dimensional $n$-th order TIs and TSCs host topologically protected edge states on their $(d-n)$-dimensional boundaries~\cite{trifunovic_higher-order_2019}. For instance, a 2D (3D) second order TI and TSC support corner (hinge) states.

A prominent example of such phases is the two-dimensional (2D) second-order TSC (SOTSC), which hosts zero-dimensional Majorana corner states (MCSs)~\cite{zhu_tunable_2018, franca_phase-tunable_2019, volpez_second-order_2019, Plekhanov_Floquet_2019, Laubscher_Fractional_2019, zhang_topological_2020, wu_-plane_2020, ghosh_floquet_2021,subhadarshini_engineering_2025}. Theoretical predictions suggest that SOTSCs offer significant experimental advantages over their first-order 1D counterparts, including the potential to exploit high-temperature superconductors~\cite{wang_high-temperature_2018, yan_majorana_2018}. There is no need to engineer additional gaps to localize MZMs into discrete positions in 2D systems. The edges are fully gapped, which enhances robustness against disorder and quasiparticle poisoning, while crystalline-symmetry-driven mass inversions reduce the need for fine‑tuning external fields or applying complicated device geometries. This reduces fabrication complexity: there is no need for fine-tuning magnetic fields, applying complicated junction geometries, or external gates. Furthermore, their planar geometry naturally facilitates nonlocal braiding protocols; rather than relying on complex electrostatic gate induced physical movements in quasi 1D Majorana networks~\cite{ alicea_non-abelian_2011, clarke_majorana_2011} or intricate gate landscapes or resonant manipulation schemes~\cite{kornich_braiding_2021}. It is expected that MCSs can be efficiently exchanged by rotating in-plane magnetic fields~\cite{zhu_tunable_2018, volpez_second-order_2019, wu_-plane_2020}. Consequently, SOTSCs constitute a highly promising experimental platform for hosting and manipulating exotic quasiparticles like MCSs for the purpose of topological quantum computation.

Despite these advantages, the experimental verification of the HOTSC phase, as well as the detection and control of the MCS states, remains a major challenge. In addition, the spin-related properties of HOTSC~\cite{plekhanov_quadrupole_2021, wang_topological_2022, lee_spinful_2023, wang_spin-dependent_2023, zhang_spin-selective_2026} are still a largely unexplored area, especially the spin resolved local and nonlocal transport characteristics of HOTSC systems, which can be experimentally probed, providing direct access to their underlying topological and superconducting properties.

In this work, we numerically investigate the spin and transport properties of 2D SOTSCs hosting a single pair of MCS. We demonstrate that MCSs in these systems exhibit a unique spatial profile of electronic spin texture, specifically in the direction perpendicular to an applied in-plane magnetic field within the $xy$-plane. Crucially, we find that the MCSs in a pair have opposite spin-polarization signs, allowing them to be labeled by their spin. To detect such spin signatures, we propose a spin-resolved quantum transport protocol based on local and nonlocal differential conductance measurements utilizing spin-polarized and normal scanning tunneling microscopy (STM) probes. Specifically, we show the local conductance signal is governed by the selective spin Andreev reflection (SSAR) mechanism~\cite{he_selective_2014, sun_majorana_2016, hu_theory_2016, maska_polarization_2017, jeon_distinguishing_2017, wang_spin-polarized_2021, zeng_spin_2025}, where the magnitude of the zero bias conductance peak (ZBCP) depends on the alignment between the spin polarization of the STM tip and the local quasiparticle spin density in a given direction. In the nonlocal transport regime, we find a direct correspondence between the sign of the nonlocal conductance and the sign of the probed MCS spin density, resulting from the competition between crossed Andreev reflection (CAR) and elastic cotunneling (ECT) based on the spin polarization of the  quasiparticles in the superconductor~\cite{melin_sign_2004, Beckmann_Evidence_2004, Trocha_Spin-resolved_2015, trocha_cross_2018, bordoloi2022spin, szumniak_spin-resolved_2024}.

In our analysis, we focus on a system with a disk-shaped geometry. This choice is directly motivated by TQC applications, as the disk configuration ensures that the topological energy gap remains open during the in-plane rotation of the magnetic field. This is a critical requirement for protecting quantum information from errors while exchanging MCSs within the pair~\cite{Park_Detecting_2015, Phong_Majorana_2017, Park_Electron_2020, Yang_Rotating_2022}. Moreover, our results show that these spin-related features remain robust in the presence of moderate static disorder, offering a clear experimental route to probe the spin structure of superconducting quasiparticles hosted in SOTSCs, making important steps towards the control and detection of MCS-based qubits.

\begin{figure}[ht!]
	\centering
	\includegraphics[width=8.6cm]{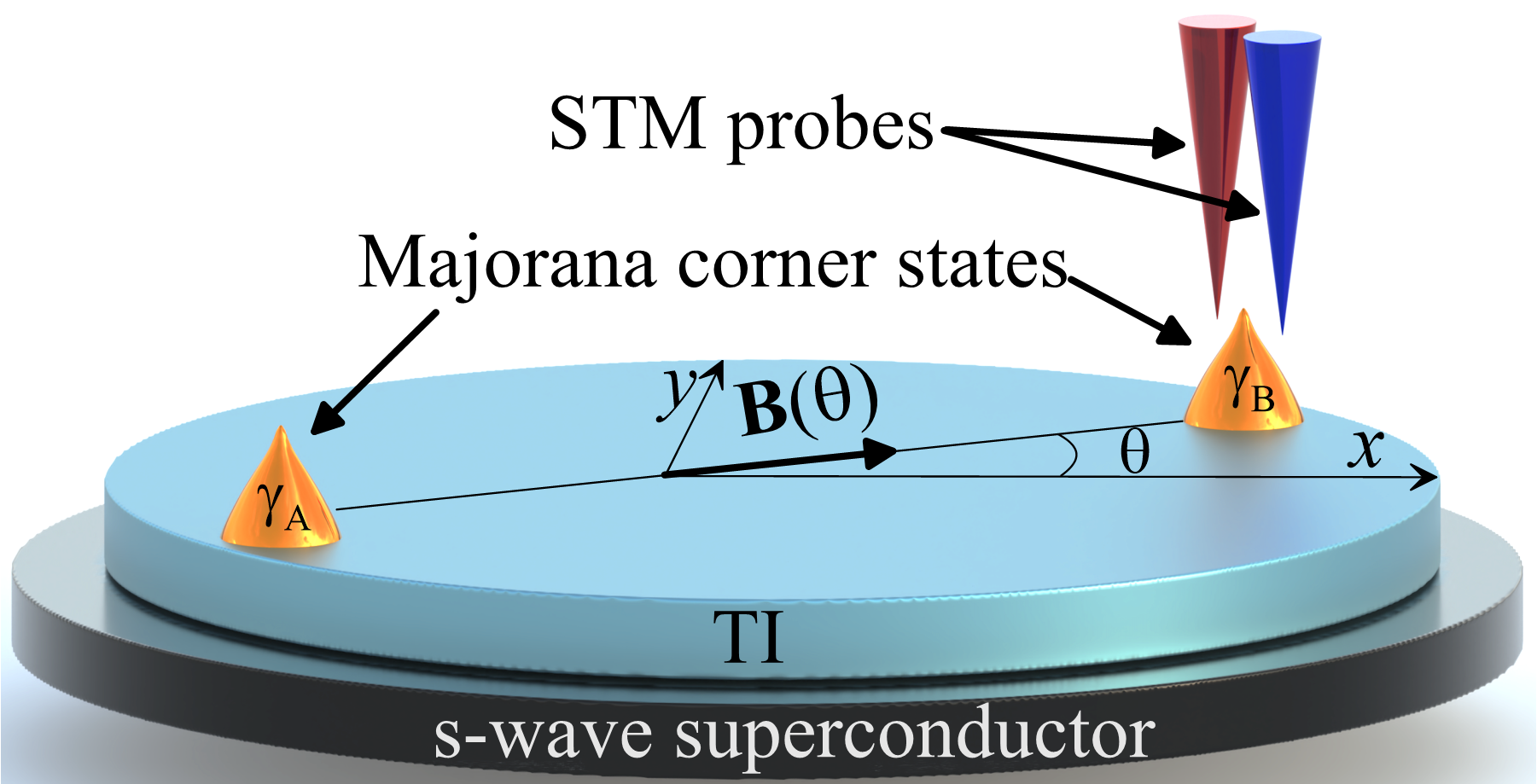}
	\caption{Schematics of a TI disk proximity-coupled to an s-wave superconductor. An external magnetic field {\bf B}($\theta$) is applied in the $xy$-plane at an angle $\theta$. In the second-order topological superconducting phase, the system hosts a pair of zero-energy MCSs (orange), which are localized along the disk boundary at positions determined by the direction of the in-plane magnetic field {\bf B}($\theta$). The STM probes (shown as vertical cones) are placed close to the MCS wave function and are employed to investigate both local and nonlocal transport properties of the studied system.}
	\label{fig:System}
\end{figure}


\section{ Model.} We investigate a two-dimensional SOTSC hosting a single pair of Majorana corner modes, whose positions can be controlled by adjusting the orientation of an in-plane magnetic field. The considered system can be implemented by placing an $s$-wave superconductor in proximity to a topological insulator while applying an external in-plane magnetic field (see Fig.~\ref{fig:System}) \cite{volpez_second-order_2019, plekhanov_quadrupole_2021}. This corresponds to a basic model of helical TSC with edge modes gapped out by the external Zeeman field.

The resulting tight-binding model momentum-space Hamiltonian, assuming periodic boundary conditions, is given by ${H}=\sum_{\bf k}\Psi_{\bf k}^\dag{\mathcal H}({\bf k}){\Psi}_{\bf k}$, where the Hamiltonian density ${\mathcal{H}({\bf k})}$ takes the following form: 
\begin{align}
{\mathcal{H}({\bf k})}=&[4t-\mu-2t(\cos(k_xa)+\cos(k_y a))]\eta_0\tau_z\sigma_0\nonumber\\
&+\Gamma \eta_x\tau_z\sigma_0 +\Delta_{sc}\eta_y\tau_y\sigma_y +\mathcal{H}_z \nonumber\\
&\hspace{1pt}+2\alpha_R\eta_z[\sin(k_xa)\tau_z\sigma_y-\sin(k_ya)\tau_0\sigma_x]\label{hamiltonian_position},
\end{align}
\noindent
with the Zeeman term $\mathcal{H}_Z$:
\begin{align}
\mathcal{H}_Z=\Delta_Z\eta_0[\cos(\theta)\tau_z\sigma_x+\sin(\theta)\tau_0\sigma_y]
\end{align}
where the basis state vector ${\Psi}_{\bf k}~=~({c}_{{\bf k} \uparrow1}, {c}_{{\bf k} \downarrow1}, {{c}^{\dag}_{{\bf k} \uparrow1}}, {{c}^{\dag}_{{\bf k} \downarrow1}},{c}_{{\bf k} \uparrow\bar{1}}, {c}_{{\bf k} \downarrow\bar{1}}, {{c}^{\dag}_{{\bf k} \uparrow\bar{1}}}, {{c}^{\dag}_{{\bf k} \downarrow\bar{1}}})^T$ is given in the standard Nambu representation. The corresponding quasiparticle wave function in real space takes the following form ${\Psi}_{\bf r}=({c}_{{\bf r} \uparrow1}, {c}_{{\bf r} \downarrow1}, {{c}^{\dag}_{{\bf r} \uparrow1}}, {{c}^{\dag}_{{\bf r} \downarrow1}},{c}_{{\bf r} \uparrow\bar{1}}, {c}_{{\bf r} \downarrow\bar{1}}, {{c}^{\dag}_{{\bf r} \uparrow\bar{1}}}, {{c}^{\dag}_{{\bf r} \downarrow\bar{1}}})^T$~(see Appendix A for details). The creation operator ${c}^{\dag}_{{\bf r} \sigma\tau}$ acts on an electron with spin (orbital) $\sigma=\uparrow,\downarrow$ ($\eta=1, \bar{1}$) located at site ${\bf r}=(ia,ja)=(x,y)$ in a square lattice with a lattice spacing $a$ used in the effective tight-binding model.
The Zeeman splitting  $\Delta_Z$ is determined by the $g$-factor and by the strength of the external inplane magnetic field ${\bf B}(\theta)$. The superconducting pairing term  $\Delta_{sc}$ is induced in the TI via the proximity effect by the $s$-wave superconductor. The chemical potential of the system $\mu$ is calculated from the SOI energy, and the hopping amplitude $t=t_x=t_y=\hbar^2/(2m^*a^2)$ is isotropic in the $x$ and $y$ directions, where $m^*$ is the effective mass. The Pauli matrices $\sigma_i$ ($\tau_i$) [$\eta_i$] act on spin (particle-hole) [e.g., electron orbital or layer index] degrees of freedom, and $\alpha$ denotes the strength of the Rashba SOI, whereas $\Gamma$ characterizes, e.g., the magnitude of the electron orbital or interlayer coupling. By diagonalizing numerically a finite size Hamiltonian ${H}$, we find the spectrum $E_n$ and corresponding wavefunctions $\Psi_\nu (i,j)$ labeled by the index $\nu=-4N+1,...,4N$, where $N$ is the number of system sites. Here, we consider a finite size system of disk geometry with a radius $R$. Throughout this work, all quantities with dimensions of energy are expressed in units of the hopping amplitude $t$. Furthermore, we fix the lattice constant to $a = 1$, such that the spatial coordinates are given by ${\bf r}=(i,j)$, which we further denote as ${\bf r}=(x,y)$ to avoid notational ambiguities with other quantities or indices.

In order to investigate the spin properties of quasiparticles hosted in the analyzed SOTSC, especially zero energy MCSs, we consider the electronic part of the spin density~\cite{sticlet_spin_2012,li_majorana_2018} in the ${\bf m}_\beta=(\cos(\beta),\sin(\beta), 0)$ direction in the $xy$-plane, which, for the $\nu$-th quasiparticle eigenstate $\Psi_\nu (\textbf{r})$, is defined as follows:
\begin{align}
S^e_{\beta}(\textbf{r}, E_\nu)=\Psi_\nu^\dag(\textbf{r})\eta_0\frac{(\tau_0+\tau_z)}{2}({\bf m}_\beta \cdot\vec{\sigma})\Psi_\nu(\textbf{r})
\end{align}
where $\vec{\sigma}$ is a vector of Pauli matrices.
In our analysis, we will further focus exclusively on components of electronic spin density in the directions parallel and perpendicular to the magnetic field {\bf B}($\theta$) in the $xy$-plane, denoted respectively as $S^e_{\beta^\parallel_\pm}(i,j)$ and $S^e_{\beta^\perp_\pm}(i,j)$, where $\beta^\parallel_\pm=(\theta+\pi/2) \mp \pi/2$ and $\beta^\perp_\pm=\theta\pm\pi/2$.

To investigate the transport properties of the system, we analyze a configuration in which either one or two STM tips are attached. We numerically calculate the matrix elements of the differential conductance, $G_{ij} = dI_i/dV_j$, defined as the derivative of the total current $I_i$ flowing from lead $i$ into the nanowire with respect to the voltage bias $V_j$ applied to lead $j$, following the procedure described in detail in Refs.~\cite{danon_nonlocal_2020,pan_three-terminal_2021,hess_local_2021}, while the superconductor is grounded. The method is based on the Blonder-Tinkham-Klapwijk formalism~\cite{blonder_transition_1982} in which the zero-temperature local (nonlocal) conductance $G_{ii}(E)$ [$G_{ij}(E)$] between leads $i$ and $j$ ($i,j=1,2$) is given by:
\begin{align}
G_{ij}(E)=\frac{e^2}{\hbar}\left[\delta_{ij}N_i+(-1)^{\delta_{ij}}(T_{ij}^{ee}(E)-T_{ij}^{he}(E))\right]
\label{Conductance_plot},
\end{align}
where $\delta_{ij}$ is the Kronecker delta, $N_i$ is the number of modes in the lead $i$, and $T^{ee}_{ij}(E)$, $T^{he}_{ij}(E)$ are, respectively, the normal (electron-electron) and Andreev (electron-hole) transmission probabilities corresponding to the situation when the electron is injected into lead $j$ and flows out of the device as an electron or hole through lead $i$. The terms $T^{ee}_{ii}(E)$ and $T^{he}_{ii}(E)$ correspond to normal and Andreev reflection. For the system studied in this work, we compute the conductance $G_{ij}(E)$ numerically using the Kwant~\cite{groth_kwant_2014} package, supported by Adaptive~\cite{Nijholt2019} package to achieve optimal sampling of the parameter space. In all results shown, the conductance is expressed in units of $e^2/h$. In our simulations, we use two types of STM tips: normal (spin-unpolarized) and spin-polarized. Adopting the approach detailed in~\cite{szumniak_spin-resolved_2024}, the leads are represented by a one-dimensional chain of sites and described by the Hamiltonian $H_{\text{lead}}(k) = \left(2t[1 - \cos(ka)] - \mu_{\rm lead}\right)\eta_0\tau_z\sigma_0 + M_\beta\eta_0\tau_z({\bf m}_\beta\vec{\sigma})$. This form reduces to a Hamiltonian for a normal tip when $M_\beta = 0$ (here we also set $\mu_{lead}\neq 0$), and represents a spin-up (spin-down) polarized lead with spin aligned parallel (antiparallel) to the ${\bf m}_\beta$ vector for $M_\beta < 0$ ($M_\beta > 0$), and the lead potential is set to $\mu_{lead}=0$. The spin quantization axis is assumed to be in the  ${\bf m}_\beta=(\cos(\beta),\sin(\beta), 0)$ direction. The leads are tunnel-coupled to the system at the positions $\mathbf{r}^{\text{tip}}_1=(x_1, y_1)$ and $\mathbf{r}^{\text{tip}}_2=(x_2, y_2)$ respectively, for the first and second tip, with a positive tip-sample tunneling strength $t_{c} (<t)$~\cite{tersoff_theory_1983} for both tips.

\begin{figure}[ht!]
	\centering
	\includegraphics[width=8.6cm]{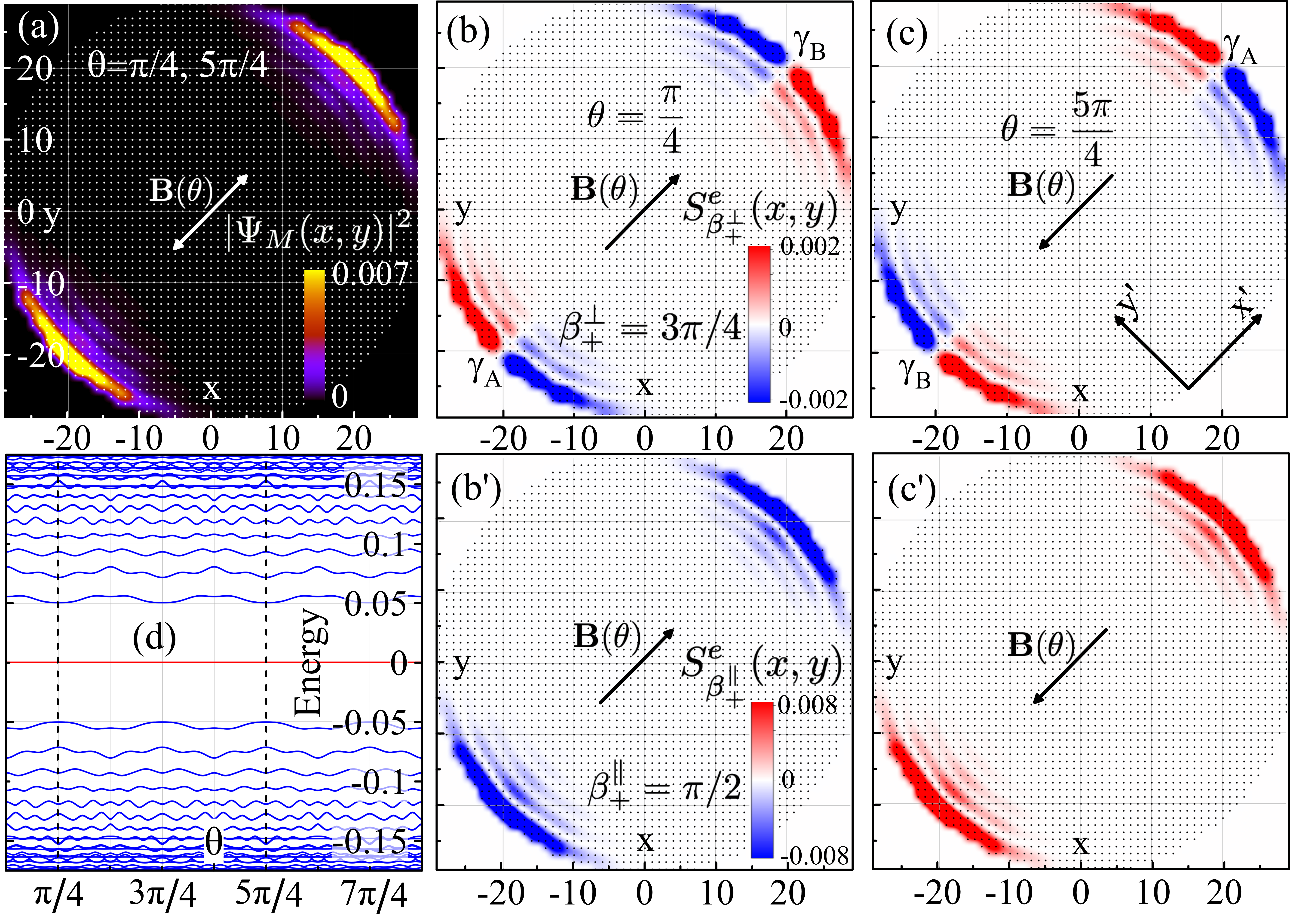}
	\caption{(a) Probability density $|\Psi_M(x,y)|^2$ of MCS for $\theta=\pi/4$ and  $5\pi/4$ hosted in SOTSC with disc geometry where system sites are marked by the dots. Electronic part of the spin density $S^e_{\beta_{+}^{\perp}}(x,y)$ of MCS in the perpendicular direction ($\beta_{+}^{\perp}=\pi/4+\pi/2$) to the in-$xy$-plane magnetic field for (b) $\theta=\pi/4$ and for (c) $\theta=5\pi/4$. We can see the quadrupolar like structure of spin polarization of the electronic part of MCS wave function. The corresponding electron part of spin density in the parallel direction $S^e_{\beta_{+}^{\parallel}}(x,y)$ ($\beta_{+}^{\parallel}=\pi/4$) is depicted on panels (b') and (c'). One can notice that when magnetic field ${\bf B}(\theta)$ is rotated by $\pi$, the position of MCSs $\gamma_A$ and $\gamma_{B}$ is interchanged. While the total probability density $|\Psi_{M}(x,y)|^2$ is not affected, we can see the sign change (reversal) of $S^e_{\theta_{+}^\perp}(x,y)$ [$S^e_{\beta_{+}^\perp}({\bf r}_{\gamma_A}) = - S^e_{\beta_{+}^\perp}({\bf r}_{\gamma_B})$], which can be signature of exchanging position of two MCS. Later we will demonstrate that this can be precisely detected by the local and particularly nonlocal transport measurement with the STM tips. In (d), we plot the energy spectrum of the system from Fig.~\ref{fig:System} as a function of the orientation of the in-plane Zeeman magnetic field $\theta$, which clearly shows presence of zero energy MCSs and gap which remains open for all values of $\theta$. Here the system parameters are set to $\Delta_Z=0.16 t$, $\Delta_{sc}=0.7t$, $\mu=0$, $\Gamma=t$, $\alpha=0.35t$ and disk radius $R=30a$.}
	\label{fig:MCS_spin_densities}
\end{figure}

\section{RESULTS}
\subsection{Spin polarization of MCS}
First, we consider a system of disk geometry and choose a set of parameters for which the system is in the SOTSC phase ($\Delta_Z<\Delta_{sc}$ and $\Gamma>|\Delta_{sc}+\Delta_Z|$), hosting a single pair of MCS~\cite{volpez_second-order_2019}. We calculate the energy spectrum $E(\theta)$ as a function of angle $\theta$, which is illustrated in Fig.~ \ref{fig:MCS_spin_densities}~(d). As a result, one can see the clear presence of a degenerate pair of zero energy MCS states (described with $\Psi_M(x,y)=\Psi_{\nu=0,1}(x,y)$ ) for all values of $\theta$. Furthermore, the energy gap remains open during the exchange operation of two MCSs, i.e., for all values of $\theta$, which is crucial for TQC, with no leakage into bulk states during the exchange. Although the bulk Hamiltonian in Eq.~(1) is rotationally invariant, the slight angular dependence of the energy gap as a function of $\theta$ can be attributed to finite-size effects resulting from the reduced rotational symmetry of the disk geometry when it is implemented on a square lattice. Next, we plot the probability density of MCSs $|\Psi_{M}(x,y)|^2$, which has the same localized shape along the edge of a disk for $\theta=\pi/4$ and  $\theta=5\pi/4$. This shows that the rotation of the magnetic field $\bf B$($\theta$) by $\pi$ does not affect the probability density of MCS. Then we present the perpendicular (parallel) components of the electronic spin density for MCS denoted as $S^e_{\beta^\perp_+}(x,y)$ [$S^e_{\beta^\parallel_+}(x,y)$], with $\beta^\perp_+=3\pi/4$ and $\beta^\parallel_+=\pi/4$ for both $\theta=\pi/4$ and $\theta=5\pi/4$ (see Fig.~\ref{fig:MCS_spin_densities}~(b-c')). For the purpose of discussing symmetry for an arbitrary angle $\theta$, we can introduce a new axis $x'$ and $y'$, respectively parallel and perpendicular to the ${\bf B(\theta)}$ field. Interestingly, one can note that for the perpendicular component of the electronic spin density and for $\theta=\pi/4$, each of the Majoranas has opposite spin polarization. To be specific, the perpendicular component of the spin density $S^e_{\beta^\perp_+}(x',y')=-S^e_{\beta^\perp_+}(-x',y')$ is reflection antisymmetric about the $y'$ axis and reflection antisymmetric about the $z'$ axis $S^e_{\beta^\perp_+}(x',y')=-S^e_{\beta^\perp_+}(x',-y')$. In the case of the parallel component of the spin density, we have $S^e_{\beta^\parallel_+}(x',y')=S^e_{\beta^\parallel_+}(-x',y')$ and [$S^e_{\beta^\parallel_+}(x,y)=S^e_{\beta^\parallel_+}(-x,-y)$], which correspond to point reflection through the origin. In both cases, the rotation of the magnetic field by $\pi$ induces a flip of the sign of spin density [see Fig.~\ref{fig:MCS_spin_densities}(c) and (c')]. However, only the spin density component that is perpendicular to the magnetic field direction differs (is opposite) between the individual MCS within a pair. This further allows us to label "MCSs" within a pair. We note that an analogous behavior of opposite spin polarization of zero energy Majorana modes has been identified in 1D TSC based on 1D proximitized Rashba nanowires~\cite{sticlet_spin_2012}. Furthermore, in Appendix B, we demonstrate that the presence of moderate onsite static disorder does not affect the electronic spin polarization of MCS.

\begin{figure}[]
	\centering
	\includegraphics[width=8.6cm]{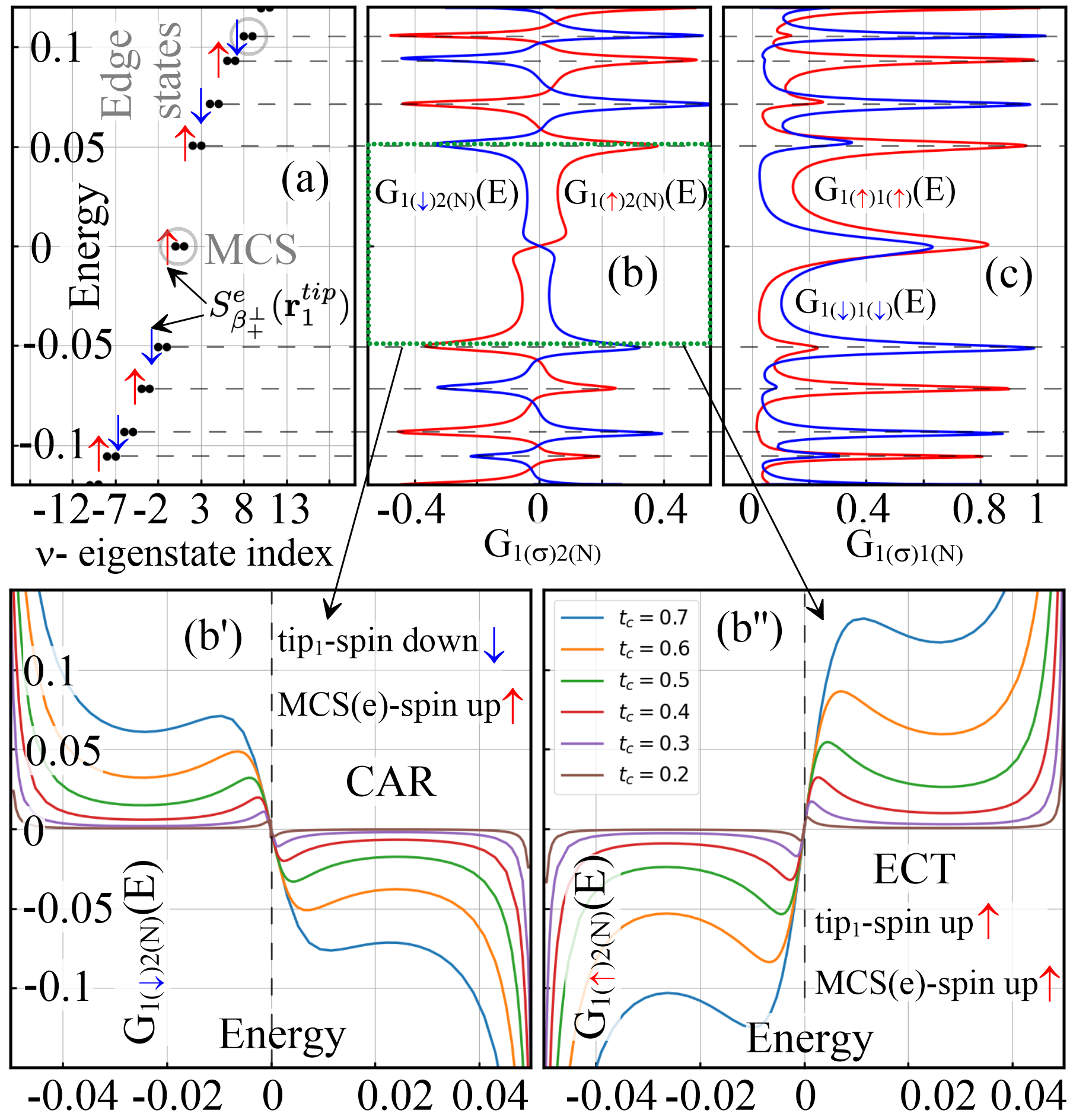}
	\caption{Part of the energy spectrum (a) of the studied system around the topological gap for $\theta=\frac{\pi}{4}$ that corresponds to the cross-section of Fig.~\ref{fig:MCS_spin_densities}(d). The panel (b) illustrates corresponding nonlocal conductance $G_{1(\sigma)2(N)}(E)$ for the setup where spin polarized STM is attached to region of MCS ($\mathbf{r}^{\text{tip}}_1=(-24a, -16a)$) with positive spin density ($S^e_{\beta_{+}^\perp}(x,y)>0$) and second normal STM tip$_2$ is attached to the area of MCS [$\mathbf{r}^{\text{tip}}_2=(-20a, -20a)$], where the MCS electronic spin density is equal zero ($S^e_{\theta_{+}^\perp}(x,y)\approx0$) while probability density has its maximum. One can clearly see (b) positive (negative) nonlocal conductance signal and (c) strong (weak) local conductance for the case when the sign of the spin polarization of the tip is the same (opposite) as the sign of local electronic spin density of probed quasiparticle state: zero energy MCS or the higher energy edge states. Lower panels represent $90^\circ$ rotated zoom of panel (b) around zero energy for negative (b') and positive (b'') spin polarization of the probe for different values of sample-tip coupling strength $t_c=0.2, 0.3,\dots ,0.7 $. One can see peaks of nonlocal conductance close to zero energy which are associated with transport through single "Majorana" from a pair. The parameters of the system are identical as in Fig.~\ref{fig:MCS_spin_densities}. The results for upper panels (b) and (c) were obtained for $t_c=0.7$. For spin polarized STM we took $|M_\beta|=0.2t$ and $\mu=0$, while for normal one we set $M_\beta=0$ and $\mu=0.2t$}
	\label{1D_nonlocal}
\end{figure}

\begin{figure}[ht!]
	\centering
	\includegraphics[width=8.6cm]{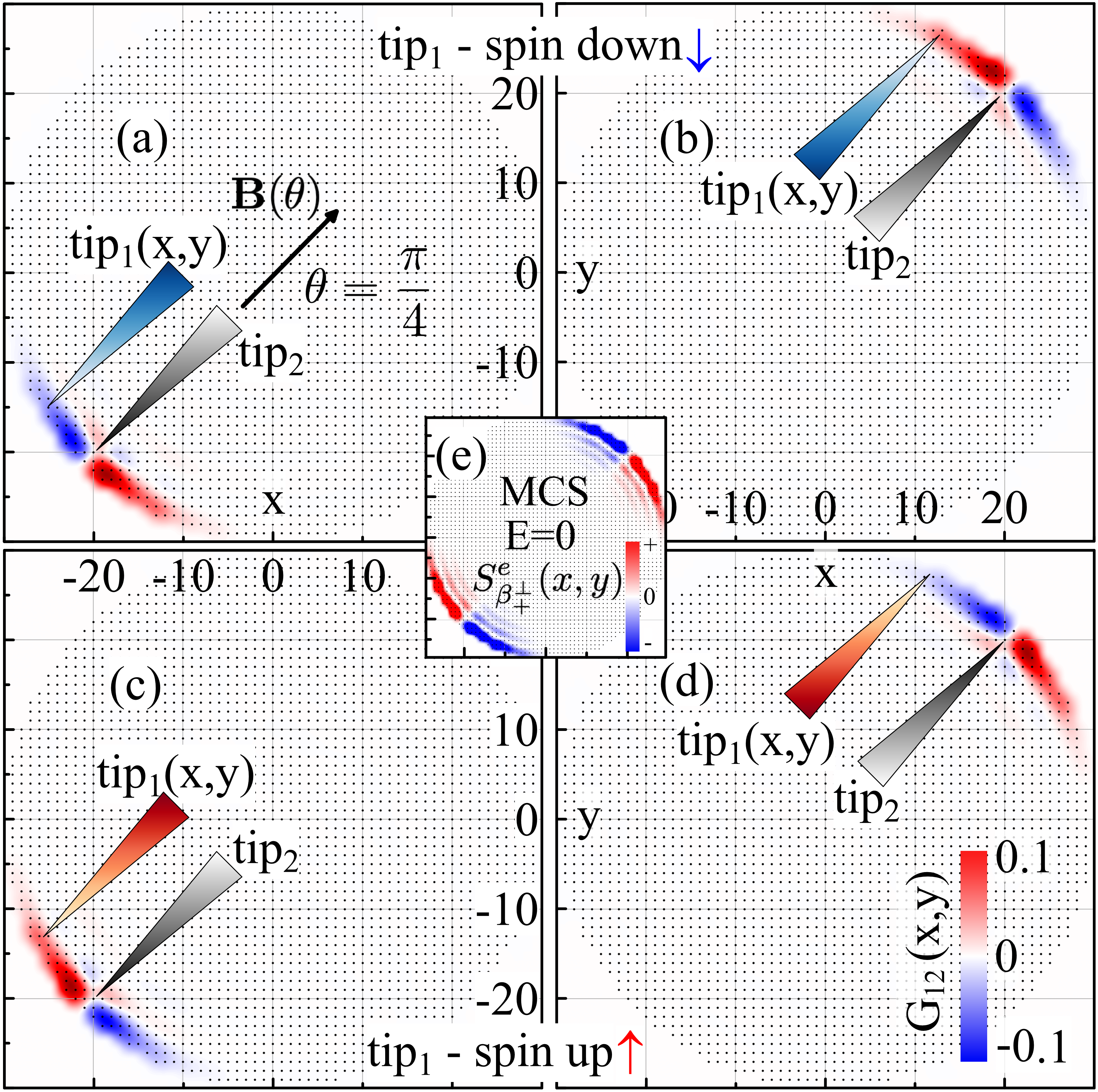}
	\caption{nonlocal conductance maps $G_{1(\sigma)2(N)}(\mathbf{r}^{\text{tip}}_1)$ as a function of position $\mathbf{r}^{\text{tip}}_1=(x, y)$ of spin polarized STM tip$_1$ which scans the surface of the system. The position of the second normal tip$_2$ is fixed to $\mathbf{r}^{\text{tip}}_2=(-20a, -20a)$ in panels (a) and (c) and to $\mathbf{r}^{\text{tip}}_2=(20a, 20a)$ for panels (b) and (d) and further marked by the gray triangle. The upper (lower) panels (a, c) [(b, d)] correspond to the situation where the spin polarization of the tip$_1$ is negative~$\sigma=\downarrow/-$ (positive~$\sigma=\uparrow/+$). Parameters are same as in Fig.~\ref{fig:MCS_spin_densities}. The coupling between STM tips is set to $t_c=0.7$ and the energy of the injected charge carries is set to $E_F\approx0.01$ corresponding to the nonlocal conductance peak/dip presented in the Fig.~\ref{1D_nonlocal}. It can be clearly seen that in such a setup nonlocal conductance $G_{1(\sigma)2(N)}(x,y)$ maps with high accuracy the electronic part of the spin density of the MCS $S_{\beta^\perp_+}^e(x,y)$ [panel (e)] from Fig.~\ref{fig:MCS_spin_densities}~(b). }
	\label{G12_xy_MCS}
\end{figure}

\begin{figure}[ht!]
	\centering
	\includegraphics[width=8.6cm]{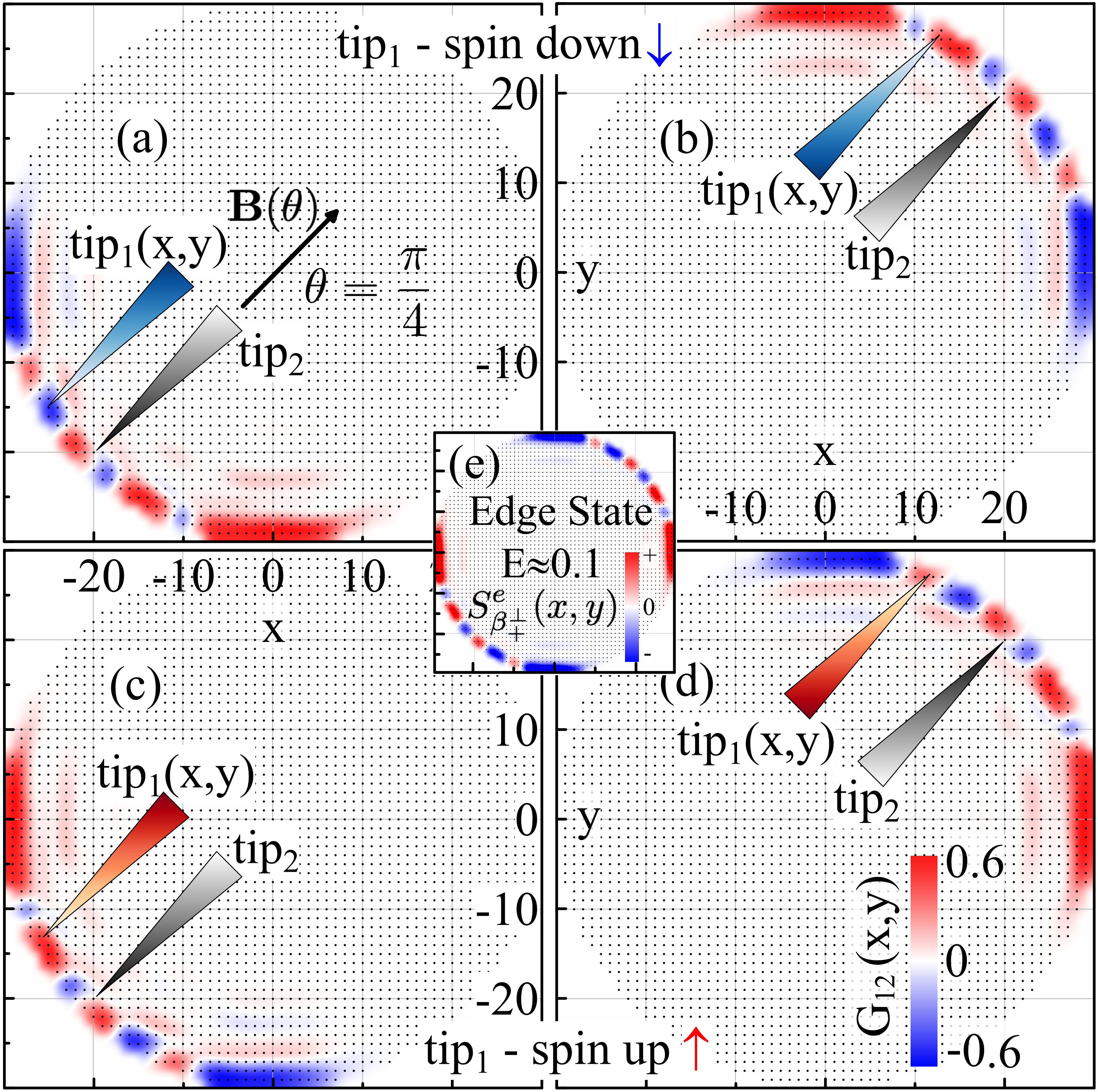}
	\caption{Same as Fig.~\ref{G12_xy_MCS} but for the energy of the charge carries in the leads set to $E\approx 0.105$ which corresponds to the energy of the edge state marked by the gray oval in Fig.~\ref{1D_nonlocal}~(a). Again one can see that the STM tip position dependent nonlocal conductance $G_{1(\sigma)2(N)}(\mathbf{r}^{\text{tip}}_1)$ maps very well the electronic spin density of the probed edge state despite its strong oscillating character as shown on the inset (e). }
	\label{G12_xy_edge}
\end{figure}

\begin{figure}[]
	\centering
	\includegraphics[width=8.6cm]{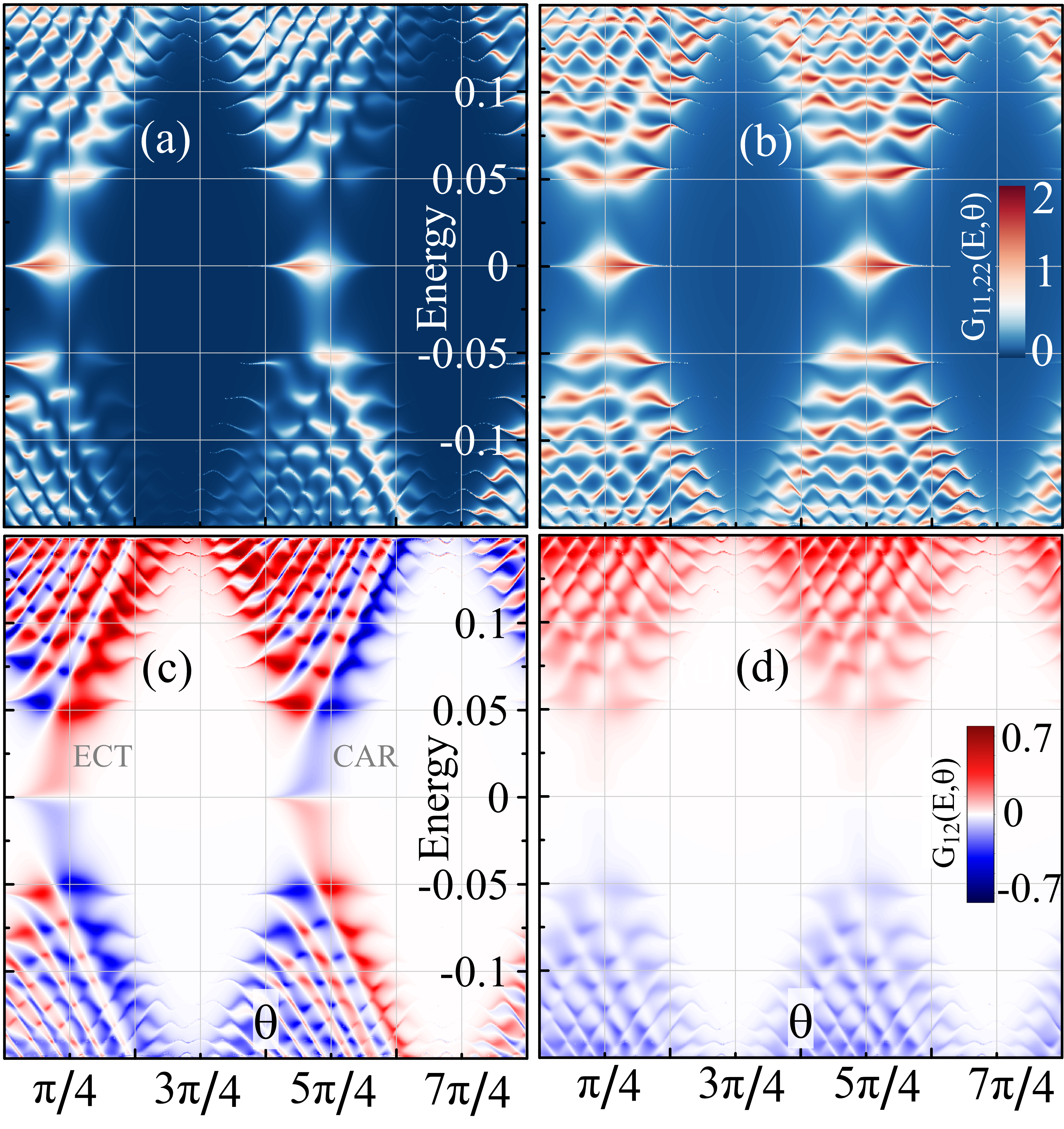}
	\caption{ Maps of local conductance (a) $G_{1(\sigma)1(\sigma)}(E,\theta)$ and (b) $G_{2(N)2(N)}(E,\theta)$ as a function of the energy of the injected charge carries and the orientation of in-plane magnetic field $\theta$ for the SOTSC with disc geometry and two STM tip probes placed at $\mathbf{r}^{\text{tip}}_1=(-24a, -16a)$ and $\mathbf{r}^{\text{tip}}_2=(-20a, -20a)$. Here, the same as before one STM tip$_1$ is spin polarized in the ${\bf m}_{\beta_{+}^\perp}~=~[\cos(3\pi/4), \sin(3\pi/4),0]$ direction ($\sigma=\uparrow/+$) while the second STM tip$_2$ is normal (spin unpolarized). Panel (c) illustrates nonlocal conductance $G_{1(\sigma)2(N)}(E,\theta)$ for the same STM tip setup, while panel (d) shows the nonlocal conductance $G_{1(N)2(N)}(E,\theta)$ for the case where two tips are normal. Local conductance measurement allows to detect zero bias signature od MCS. There are regions of $\theta$ for which ZBP disappears and this is due to the fact that for those $\theta$ angles MCS wave function does not overlap the position of the STM tip. In the case of spin polarized tip$_1$, we can also see that the ZBP $G_{1(\sigma)1(\sigma)}(E,\theta)$ signal is stronger for the $\theta\approx\pi/4$ than for $\theta\approx5\pi/4$. In first case the polarization of the tip$_1$ is the same as in the MCS while in the second case it is opposite and can be explained by the local SSAR effect and is consistent with results presented in Fig.~\ref{Conductance_plot}. This behavior is also in agreement with the nonlocal conductance $(G_{1(\sigma)2(N)}(E,\theta)$ results presented in panel (c). Because this quantity changes sign when the MCS spin polarization is reversed during the Zeeman-field-driven exchange of MCS, such a configuration can be used to identify the exchange of MCS. Furthermore, in the case where both tips are normal, shown in panel (d), we observe no sign change in the nonlocal conductance and no corresponding feature in the local conductance in panel (b), as can be expected.}
	\label{Fig_exchange}
\end{figure}

\subsection{Transport detection of quasiparticle spin density}
Next, we show that the sign of the electronic part of quasiparticle spin for MCS, as well as for arbitrary  higher energy states (e.g., edge states), can be detected both in local and nonlocal transport with appropriately chosen spin polarization of one of the STM tips, e.g., tip$_1$, while keeping the second tip$_2$ normal (spin-unpolarized). Here we choose a set of parameters corresponding to those from Fig.~\ref{fig:MCS_spin_densities}(b) and set the position of the spin polarized tip, e.g., to $\mathbf{r}^{\text{tip}}_1=(-24a, -16a)$ and the second normal tip to $\mathbf{r}^{\text{tip}}_2=(-20a, -20a)$. For such parameters, we calculate the energy spectrum of the system as shown in Fig.~\ref{1D_nonlocal}(a), which corresponds to the cut of Fig.~\ref{fig:MCS_spin_densities}(d) for $\theta=\pi/4$.  The local spin polarization of the electronic contribution to the MCS at the position of tip$_1$~-~$S^e_{\beta_{+}^\perp}(\mathbf{r}^{\text{tip}}_1)$, together with the spin polarization of tip$_1$ itself, when aligned parallel (antiparallel) to the direction of ${\bf m}_{\beta_{+}^\perp}$, is represented schematically by an arrow or sign, $\sigma = \uparrow / +$ ($\sigma = \downarrow / -$). Here, the angle $\beta_{+}^\perp$ is defined as $\beta_{+}^\perp = \pi/4 + \pi/2$.

Then we calculate the local $G_{1(\sigma)1(\sigma)}(E)$ and nonlocal $G_{1(\sigma)2(N)}$ differential conductance. It can be clearly seen that if the spin of the tip$_1$ matches (is opposite to) the local electronic spin density of the probed state, there is a strong positive (negative) nonlocal conductance $G_{1(\sigma)2(N)(E)}$ signal, which is in agreement with the theoretical expectations~\cite{szumniak_spin-resolved_2024}. Complementarily, the magnitude of the local conductance $G_{1(\sigma)1(\sigma)}(E)$ is stronger (weaker) if the spin polarization of the STM probe is aligned (antialigned) with the local spin of the system. This is consistent with the SSAR effect~\cite{he_selective_2014, sun_majorana_2016, wang_spin-polarized_2021}. Interestingly, one can notice a pair of nonlocal conductance peaks (positive and negative) around zero energy resulting from the nonlocal transport through the single MCS from the pair, where the sign again depends on the alignment or antialignment between the spin polarization of the system and the tip. We note that similar nonlocal conductance peaks around zero energy have been theoretically reported for systems hosting Majorana vortex modes~\cite{sbierski_identifying_2022}, although in that work, the study focused on spin-unpolarized transport with normal leads. 

We now demonstrate that this effect can be used as a sensitive probe of the spin density associated with quasiparticle states, for both zero-energy MCS and finite-energy excitations (e.g., edge states), through measurements of nonlocal conductance. To be specific, in such a setup, the position of the normal tip$_2$ is fixed $\mathbf{r}^{\text{tip}}_2=(-20, -20a)$ and the spin polarized tip$_1$ is scanning the surface of the probed system $\mathbf{r}^{\text{tip}}_1=(x,y)$. The corresponding nonlocal conductance results for zero energy MCS (higher energy edge state) are presented in Fig.~\ref{G12_xy_MCS} [Fig.~\ref{G12_xy_edge}]. We considered a case where the STM tip$_1$ is either spin-down or spin-up polarized, and the normal tip is attached near the maximum probability density of either "$\gamma_A$" or "$\gamma_B$" MCS, respectively, e.g., $\mathbf{r}^{\text{tip}}_2=(-20a,-20a)$ or $\mathbf{r}^{\text{tip}}_2=(20a,20a)$. The inset presents the electronic part of the spin density $S_{\beta_+^\perp}^e(x,y)$ of the probed quasiparticle state for direct comparison. One can clearly see a very good correspondence between the sign and magnitude of the quasiparticle spin density and the sign of the nonlocal conductance map $G_{1(\sigma)2(N)}$, obtained while the spin‑polarized tip$_1$, located at $\mathbf{r}^{\text{tip}}_1 = (x,y)$, scans the surface of the system, which can be written as:
\begin{align}
\text{sgn}[G_{1(\sigma)2(N)}(E,\mathbf{r}^{\text{tip}}_1)]\approx\sigma \text{sgn}[S^e_{\beta^\perp_{+}}(E,\mathbf{r}^{\text{tip}}_1)].
\end{align}
Overall, we observe a highly precise correspondence between the sign of the electronic quasiparticle spin density and the nonlocal conductance. Notably, this technique successfully captures even the pronounced spatial variations in the spin profiles of the edge states. Thus, the proposed method can be regarded as a reliable technique for the tomographic reconstruction~\cite{chevallier_tomography_2016} of the quasiparticle spin density via spin-resolved nonlocal STM transport measurements. We note that experimental realization may become highly challenging when the STM tips are brought into close proximity. However, our results remain valid over a wide range of inter-tip distances, which should facilitate experimental implementation, at least for specific subregions of the sample. An alternative strategy would be to attach a side or bottom normal contact instead of tip$_2$, analogous to the setup proposed in Ref.~\cite{sbierski_identifying_2022} for probing vortex states with a pair of normal leads.

\subsection{Detection of MCS exchange}
Finally, we propose a setup where the spin polarization can be used to detect the exchange of an MCS pair, utilizing the local and nonlocal transport techniques discussed in the previous section. Here, we suggest an experimentally feasible approach in which the positions of the two STM probes remain fixed. We use a tip configuration similar to our previous setup: a spin-polarized probe, $\text{tip}_1$, aligned along the $\mathbf{m}_{\beta_{+}^\perp}$ ($\beta_{+}^\perp = \pi/4 + \pi/2$) direction and placed at $\mathbf{r}^{\text{tip}}_1 = (-24a, -16a)$. The second probe, $\text{tip}_2$, is normal (spin-unpolarized) and fixed at the position $\mathbf{r}^{\text{tip}}_2 = (-20a, -20a)$.In this configuration, we change the orientation of the magnetic field in the $xy$-plane, which drives the exchange of the MCSs. As previously mentioned, the topological energy gap remains open for all values of $\theta$, as presented in Fig.~\ref{fig:MCS_spin_densities}(d). For this setup, we calculate the local differential conductances $G_{1(\beta_{+}^\perp)1(\beta_{+}^\perp)}(E, \theta)$ and $G_{2(N)2(N)}(E, \theta)$, as well as the nonlocal conductance $G_{1(\beta_{+}^\perp)2(N)}(E, \theta)$. For comparison, we also compute the nonlocal conductance for a setup where both leads are normal, $G_{1(N)2(N)}(E, \theta)$.The results, presented in Fig.~\ref{Fig_exchange}(a), show that around zero energy ($E \approx 0$), when the local electronic spin density of the MCS aligns with the spin polarization of the probe ($\theta \approx \pi/4$), the local conductance signal $G_{11}(E \approx 0, \theta \approx \pi/4)$ is significantly stronger than for $\theta \approx 5\pi/4$, where the MCS possesses an opposite spin polarization.However, the most striking signature of this exchange-induced spin reversal emerges in the nonlocal conductance measurement. Here, $G_{12}$ clearly changes sign from positive (for $\theta \approx \pi/4$ and $E \gtrsim 0$) to negative (for $\theta \approx 5\pi/4$ and $E \gtrsim 0$). This result holds even in the presence of strong static onsite disorder (see Appendix B for details). By contrast, in the setup utilizing two normal STM tips, the zero-bias local conductance peaks $G_{2(N)2(N)}$ are identical for both MCS spin polarizations ($\theta = \pi/4$ and $\theta = 5\pi/4$). While such a dual-normal setup is highly suitable for probing the quasiparticle charge structure of states—such as those in Rashba nanowires~\cite{szumniak_spin-resolved_2024} or 2D vortex states~\cite{sbierski_identifying_2022}—it is insensitive to the spin reversal. Finally, we note that while we have chosen a specific position for the STM tips, these transport signatures remain robust for a wide range of tip positions, provided the polarized probe sufficiently covers the local spin density of the MCS.

\section{Conclusion and outlook}

In summary, we investigated the spin polarization of superconducting quasiparticles hosted in 2D SOTSCs. We found that in the considered model, a single pair of zero-energy MCSs exhibits a distinct spin polarization perpendicular to the applied in-plane magnetic field, which is opposite for each MCS in the pair. To detect these unique spin textures, we propose a quantum transport measurement technique utilizing local and nonlocal transport via two STM probes. Specifically, our setup employs one normal and one spin-polarized STM tip, enabling a direct mapping between the sign of the electronic component of the local quasiparticle spin density and the sign (and magnitude) of the nonlocal (and local) differential conductance. This approach further allows for probing the spatial profile of the electronic spin texture of not only MCSs but also higher-energy edge states. We numerically demonstrated that this effect can be utilized to detect the exchange of an MCS pair induced by the rotation of the in-plane magnetic field, which manifests as a clear sign reversal in the measured spin polarization. Moreover, our calculations reveal that the spin-sensitive local, and particularly the nonlocal, differential conductance is highly robust against moderate static disorder.

While implementing such a setup remains experimentally challenging~\cite{nakayama_development_2012}, it is encouraging that two- and multi-probe STM transport measurements have already been successfully realized in other planar systems, such as graphene~\cite{li_electron_2013, clark_spatially_2013, clark_energy_2014}, as well as in nanowires~\cite{kolmer_two-probe_2017}.

Because the presented results demonstrate a high potential for measuring the spatial profiles of complex quasiparticle spin textures, we anticipate that this spin-resolved nonlocal detection scheme will also be suitable for probing other planar systems hosting spin-polarized quasiparticles. These include Yu-Shiba-Rusinov (YSR) states~\cite{luh1965bound, shiba1968classical, rusinov1969superconductivity, yazdani1997probing}, Caroli-de Gennes-Matricon states~\cite{caroli1964bound, Berthod_Observation_2017, chen2018discrete}, and Majorana vortex states~\cite{sun_majorana_2016, volovik1999fermion, Xu_Experimental_2015, chiu2020scalable}.

\subsection{Acknowledgements}
This work was supported within the POIR.04.04.00-00-5CE6/18 project, carried out within the HOMING programme of the Foundation for Polish Science, co-financed by the European Union under the European Regional Development Fund and the Swiss National Science Foundation. We gratefully acknowledge Polish high-performance computing infrastructure PLGrid (HPC Center: ACK Cyfronet AGH) for providing computer facilities and support within computational grant no. PLG/2024/016986. This project received funding from the European Union's Horizon 2020 research and innovation program (ERC Starting Grant, Grant No 757725). This work was supported as a part of NCCR SPIN, a National Centre of Competence in Research, funded by the Swiss National Science Foundation (Grant No. 225153). DL acknowledges the Deanship of Research  and the Quantum Center at KFUPM for the support received under Grant no. CUP25102 and INQC2600.

\appendix

\section{Tight-binding Hamiltonian}
In order to study the properties of a finite size system, one needs to transform the total momentum-space tight-binding Hamiltonian $H$ [see Eq. (1) in the main text] into its coordinate-space representation $\bar{H}$, which takes the following form on the square lattice:

\begin{align}
\bar{H}=&\sum_{i,j}\Psi^{\dag}_{i,j}(\mathcal{H}_\mu+\mathcal{H}_\Gamma+\mathcal{H}_{sc}+\mathcal{H}_Z)\Psi_{i,j} \nonumber\\
+&\sum_{i,j}[\Psi^{\dag}_{i+1,j}(\mathcal{H}_x+\mathcal{H}_x^{SO})\Psi_{i,j}+\text{H.c.}] \nonumber\\
+&\sum_{i,j}[\Psi^{\dag}_{i,j+1}(\mathcal{H}_y+\mathcal{H}_y^{SO})\Psi_{i,j}+\text{H.c.}]
\end{align}
where the Hamiltonian densities take the following forms:
\begin{align}
&\mathcal{H}_\mu= (4t-\mu_{i,j})\eta_0\tau_z\sigma_0 \\
&\mathcal{H}_\Gamma= \Gamma\eta_0\tau_z\sigma_0\\
&\mathcal{H}_{sc}= \Delta_{sc}\eta_y\tau_y\sigma_y\\
&\mathcal{H}_Z= \Delta_Z\eta_0[\cos(\theta)\tau_z\sigma_x+\sin(\theta)\tau_0\sigma_y]\\
&\mathcal{H}_x= \mathcal{H}_y= -t \eta_0\tau_z\sigma_0\\
&\mathcal{H}_x^{SO}=-i\alpha_R\eta_z\tau_z\sigma_y\\
&\mathcal{H}_y^{SO}=i\alpha_R\eta_z\tau_0\sigma_x
\end{align}
with the basis state vector in Nambu space 
${\Psi}_{\bf r}=({c}_{{\bf r} \uparrow1}, {c}_{{\bf r} \downarrow1}, {{c}^{\dag}_{{\bf r} \uparrow1}}, {{c}^{\dag}_{{\bf r} \downarrow1}},{c}_{{\bf r} \uparrow\bar{1}}, {c}_{{\bf r} \downarrow\bar{1}}, {{c}^{\dag}_{{\bf r} \uparrow\bar{1}}}, {{c}^{\dag}_{{\bf r} \downarrow\bar{1}}})^T$ and ${\bf r}=(i,j)$.
Since we work on a square lattice with disk-like boundaries, the values of indices (i, j) are bound by the disk radius R: $(ia)^2-(ja)^2\leq R$

\section{Disorder}
\begin{figure}[hb!]
	\centering
	\includegraphics[width=8.6cm]{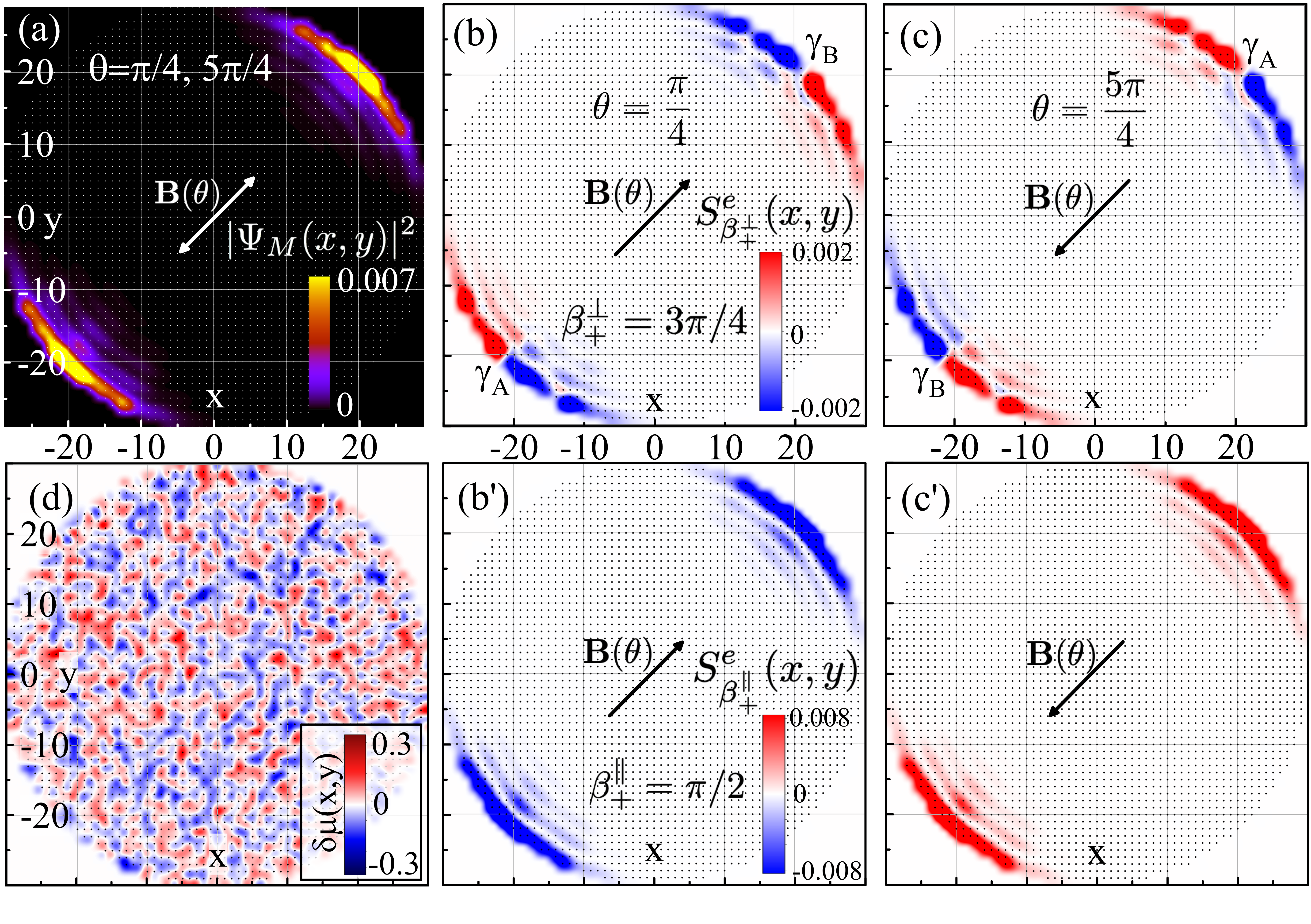}
	\caption{The same as Fig.~\ref{fig:MCS_spin_densities}$(a-c')$ but for the system with onsite potential disorder $|\delta\mu(x,y)|\lesssim 0.3$ which is illustrated on panel (d).}
	\label{Fig1_dis}
\end{figure}
\begin{figure}[]
	\centering
	\includegraphics[width=8.6cm]{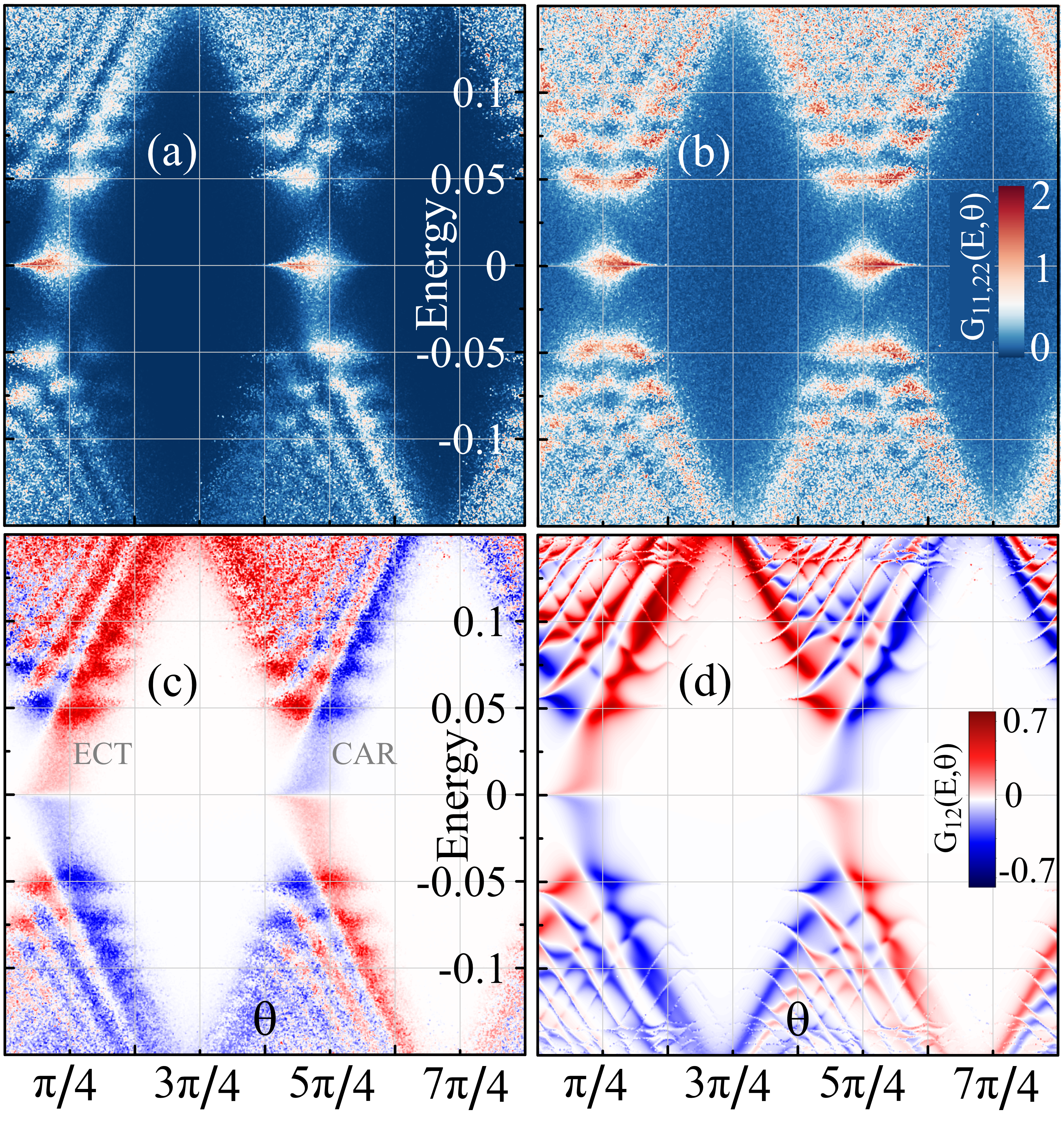}
	\caption{ Same as Fig.~\ref{Fig_exchange}(a–c), but with the three times stronger onsite disorder potential ($\sigma\mu=0.3$, $|\delta\mu(x,y)|\lesssim 1$) chosen independently for each pair $(\Delta_Z, E)$, and (d) where $G_{1,2}(\Delta_Z, E)$ is computed using the same set of disorder realizations for all parameter pairs $(\Delta_Z, E)$.}
	\label{Fig2_dis}
\end{figure}

Finally, we examine whether our main conclusions remain valid in the presence of disorder. Specifically, we model on-site fluctuations by adding random variations $\delta\mu_{i,j}$ [also denoted as $\delta\mu(x, y)=\delta\mu(ia, ja)$] to the chemical potential at each lattice site $(i,j)$, so that $\mu_{i,j} = \mu + \delta\mu_{i,j}$. The random variable $\delta\mu_{i,j}$ is taken according to a Gaussian distribution with zero mean and standard deviation $\sigma_\mu$. 

We first introduce a moderate amount of disorder by setting~$\sigma_\mu=0.1t$, which leads to typical fluctuations $|\delta\mu_{i,j}|\lesssim 0.3t$, as illustrated in Fig.~\ref{Fig1_dis}(d). In this regime, both the probability density and the spin densities are only weakly influenced by the fluctuations in the chemical potential. We then increase the disorder strength by a factor of three, setting $\sigma_\mu=0.3t$, which yields $|\delta\mu_{i,j}|\lesssim 1t$. Under these conditions, the spin densities associated with the MCSs are strongly modified; nevertheless, the underlying symmetry is preserved, and the spin-reversal nonlocal conductance signature is clearly visible [see Fig.~\ref{Fig2_dis}(c,d)], whereas the local SSAR-related transport signature becomes barely visible. 

In our quantum transport analysis, we implement the disorder potential in two distinct ways: (i) $\delta\mu_{i,j}$ is independently sampled for each pair $(\Delta_Z, E)$, as in Figs.~\ref{Fig2_dis}(a-c); (ii) a single realization of $\delta\mu_{i,j}$ is fixed and used for all values of $(\Delta_Z, E)$. In both scenarios, the nonlocal conductance signatures remain remarkably robust against disorder.

\bibliography{references}

\end{document}